
%
%
%

\magnification=\magstep1

%
%

{

\nopagenumbers

This file contains the title page, abstract, table of contents,
introductory chapter, concluding chapter, and references for my thesis.
The complete text and figures are available via anonymous ftp from
astro.princeton.edu in the directory /summers/thesis. The files are
compressed Postscript and total nearly 20 MB compressed and about 65 MB
uncompressed. Most of the size is due to large scatter plots, up to 3 MB
per plot. Please see the README file first.

\bigskip

Some problems may arise when TeX'ing this file because of the use of
several ornamental fonts. Their inclusion is only aesthetic, and are not
used in the important parts.

\bigskip

Please report any problems to me at summers@astro.princeton.edu.

\vfill\eject

}

%
%

{

\voffset=-0.02truein
\vsize=8.8truein

\baselineskip=13pt

\footline={\hfil}

\def\oneskip{\vskip12pt plus 12pt}
\def\twoskip{\vskip24pt plus 24pt}
\def\threeskip{\vskip8pt plus 8pt}
\def\fourskip{\vskip6pt plus 6pt}

\font\eightrm=cmr8

%
%

\global\pageno=-1
\null
\centerline{Cosmological Simulations of Galaxy Formation}
\centerline{Including Hydrodynamics}
\oneskip
\centerline{by}
\oneskip
\centerline{Francis Joseph Summers}
\oneskip
\centerline{B.S. (Virginia Polytechnic Institute \& State University)
1988}
\centerline{M.A. (University of California at Berkeley) 1990}
\twoskip
\centerline{A dissertation submitted in partial satisfaction}
\centerline{of the requirements for the degree of}
\oneskip
\centerline{Doctor of Philosophy}
\threeskip
\centerline{in}
\threeskip
\centerline{Astronomy}
\threeskip
\centerline{in the}
\threeskip
\centerline{Graduate Division}
\threeskip
\centerline{of the}
\threeskip
\centerline{University of California at Berkeley}
\twoskip
\leftline{\hskip 6em Committee in Charge}
\fourskip
\leftline{\hskip 8em Professor Marc Davis, Chair}
\leftline{\hskip 8em Professor Joseph Silk}
\leftline{\hskip 8em Professor Christopher McKee}
\twoskip
\centerline{1993}
\eject

%
%

{

\hoffset=0.5truein
\hsize=5.9truein

\parskip=0pt plus 3pt
\parindent=2em
\baselineskip=24pt

\raggedbottom

\vbox{
\baselineskip=13pt
\centerline{Abstract}
\oneskip
\centerline{Cosmological Simulations of Galaxy Formation}
\centerline{Including Hydrodynamics}
\oneskip
\centerline{by}
\oneskip
\centerline{Francis Joseph Summers}
\oneskip
\centerline{Doctor of Philosophy in Astronomy}
\oneskip
\centerline{University of California at Berkeley}
\oneskip
\centerline{Professor Marc Davis, Chair}
\twoskip
\oneskip
}

The formation of galaxies in hierarchical cosmogonies is studied using
high resolution N-body simulations. The equations of gravity and
hydrodynamics are evolved for two particle fluids: a pressureless dark
matter component and a dissipational baryonic component. The collapse of
structure is followed self-consistently from Mpc scale filamentary
structures to kpc scale galactic objects.

The simulated galaxy population has sizes, masses, and abundances in the
range of those for real galaxies. A large fraction of these objects show
rotationally supported disk structures. This is the first demonstration
that disk formation is a natural consequence of cosmological collapse.
Isolated galaxies generally grow through steady mass accretion while
objects in clustered regions are more likely to undergo mergers.
Galactic disks form via discrete accretion of relatively large gas
clouds and viscous transport of angular momentum to the inner regions.
The formation era is consistent with high redshift observations and has
a peak rate near redshift 3.

Galaxies collapse as segments of filaments and then flow along the
filaments toward intersections. The formation of groups and clusters of
galaxies is dominated by radial, directed infall which will increase the
relaxation timescales as well as the incidence of merging, tidal
disruption, and gas stripping. In this specific region, biases in both
the spatial and kinematic distributions of galaxies relative to the dark
matter are found. Virial mass estimates give results a factor of two to
three lower than the true value. However, finer modelling of the stellar
fluid within galaxies will be required before these findings can be
generalized.

A didactic discussion of smoothed particle hydrodynamics (SPH) is
presented along with several computational tests. Density estimation is
shown to be sensitive to the resolution of SPH parameter choices and can
greatly affect the physical results. Initial work on the self-similar
spherical accretion problem indicates the general method is sound, but
further work is required to search for optimal parameter sets. Tests of
several galaxy identification schemes within simulations show that SPH
methods are necessary for tracing galaxies through clustering and that
the phenomenon of velocity bias has not yet been convincingly
demonstrated.

\vbox{
\baselineskip=13pt
\oneskip
\twoskip
\twoskip
\rightline{\hbox to 4.0truein{\hrulefill}}
\rightline{\hbox to 4.0truein{\eightrm\quad Chair \hfill Date \qquad}}
}

\vfill\eject

}

%
%

{\leftskip=0.5truein

\def\tocchapterentry#1#2#3{\bigbreak\line{\hskip0.5truein{\bf #1}
    \hskip 0.6em{\bf #2}\hfil #3}\smallskip}
\def\tocsectionentry#1#2#3{\line{\hskip0.5truein\hskip 2em{#1}
    \hskip 0.6em{#2}\quad\dotfill\quad #3}}
\def\tocsubsectionentry#1#2#3{\line{\hskip0.5truein\hskip 4em{#1}
    \hskip 0.6em{\sl #2}\quad\dotfill\quad #3}}

\centerline{\bf Table of Contents}
\bigskip

\tocchapterentry {1} {Introduction} {1}
\tocsectionentry {1.1} {Galaxies} {2}
\tocsectionentry {1.2} {Cosmology} {4}
\tocsectionentry {1.3} {Galaxies within Cosmology} {7}
\tocchapterentry {2} {Computational Methodology} {9}
\tocsectionentry {2.1} {The Density Field of a Particle Distribution} {10}
\tocsectionentry {2.2} {P3MSPH Overview} {12}
\tocsectionentry {2.3} {P3M} {13}
\tocsectionentry {2.4} {SPH} {15}
\tocsectionentry {2.5} {Variations on the SPH Theme} {17}
\tocsectionentry {2.6} {Combining P3M and SPH} {18}
\tocsectionentry {2.7} {Calculation Details} {20}
\tocchapterentry {3} {Simulation Description and Overview} {24}
\tocsectionentry {3.1} {Simulation Parameters} {25}
\tocsectionentry {3.2} {Simulation Limitations} {27}
\tocsectionentry {3.3} {Simulation Structure on Large Scales} {29}
\tocsectionentry {3.4} {Phases of the Particle Fluids} {31}
\tocsectionentry {3.5} {Resolution and Cooling} {33}
\tocchapterentry {4} {The Simulated Galaxy Population} {45}
\tocsectionentry {4.1} {Galaxy-like Objects and Dark Matter Halos} {46}
\tocsectionentry {4.2} {Abundance and Distribution} {47}
\tocsectionentry {4.3} {Morphology} {51}
\tocsectionentry {4.4} {Galactic Disks} {54}
\tocchapterentry {5} {Galaxy Formation} {75}
\tocsectionentry {5.1} {Galaxy Formation History} {76}
\tocsectionentry {5.2} {Disk Formation and Hierarchical Collapse} {78}
\tocsectionentry {5.3} {Correspondence of Globs and Halos} {81}
\tocsectionentry {5.4} {Merging} {83}
\tocchapterentry {6} {Group and Cluster Formation} {101}
\tocsectionentry {6.1} {Hierarchical Group Formation} {101}
\tocsectionentry {6.2} {Baryonic and Dark Matter Distributions} {104}
\tocsectionentry {6.3} {Correlation Properties and Biasing} {106}
\tocsectionentry {6.4} {Estimates of Group Mass and $\Omega $} {110}
\tocchapterentry {7} {Simulations with Star Formation} {130}
\tocsectionentry {7.1} {Methods} {131}
\tocsectionentry {7.2} {Analysis} {134}
\tocsubsectionentry {7.2.1} {Star Particles and Galaxy-like Objects} {134}
\tocsubsectionentry {7.2.2} {Star Particles and Galaxy Formation} {136}
\tocsubsectionentry {7.2.3} {Star Particles and Group Formation} {139}
\tocsubsectionentry {7.2.4} {Discussion} {141}
\tocchapterentry {8} {Galaxy Tracers} {171}
\tocsectionentry {8.1} {Criteria and Methods} {172}
\tocsectionentry {8.2} {Collisionless Algorithms} {174}
\tocsubsectionentry {8.2.1} {Halos} {174}
\tocsubsectionentry {8.2.2} {Peaks} {175}
\tocsubsectionentry {8.2.3} {Couchman and Carlberg's Method} {176}
\tocsubsectionentry {8.2.4} {The Most Bound Algorithm} {178}
\tocsectionentry {8.3} {SPH Galaxy Tracers} {180}
\tocsectionentry {8.4} {Statistical Measures and Velocity Bias} {182}
\tocsectionentry {8.5} {Discussion} {184}
\tocchapterentry {9} {SPH Tests} {202}
\tocsectionentry {9.1} {Density Fields} {203}
\tocsubsectionentry {9.1.1} {Fixed Smoothing Lengths} {205}
\tocsubsectionentry {9.1.2} {Iterated Smoothing Lengths} {206}
\tocsubsectionentry {9.1.3} {SDE and KHW} {207}
\tocsubsectionentry {9.1.4} {Discussion} {208}
\tocsectionentry {9.2} {Self-similar Secondary Infall} {211}
\tocsubsectionentry {9.2.1} {Top Hat Collapse of a Collisional Gas} {213}
\tocsubsectionentry {9.2.2} {Self-similar Behavior} {214}
\tocsubsectionentry {9.2.3} {Variation of Parameters} {215}
\tocsubsectionentry {9.2.4} {Discussion} {218}
\tocchapterentry {10} {Summary and Discussion} {245}
\tocsectionentry {10.1} {Summary} {246}
\tocsubsectionentry {10.1.1} {Chapter Two} {246}
\tocsubsectionentry {10.1.2} {Chapter Three} {246}
\tocsubsectionentry {10.1.3} {Chapter Four} {247}
\tocsubsectionentry {10.1.4} {Chapter Five} {248}
\tocsubsectionentry {10.1.5} {Chapter Six} {249}
\tocsubsectionentry {10.1.6} {Chapter Seven} {250}
\tocsubsectionentry {10.1.7} {Chapter Eight} {251}
\tocsubsectionentry {10.1.8} {Chapter Nine} {252}
\tocchapterentry {10} {Summary and Discussion} {}
\tocsectionentry {10.2} {Discussion} {253}
\tocsubsectionentry {10.2.1} {Methods} {253}
\tocsubsectionentry {10.2.2} {Resolution} {254}
\tocsubsectionentry {10.2.3} {Physics} {255}
\tocsectionentry {10.3} {Future Paths} {256}
\tocsectionentry {10.4} {Some Final Remarks} {258}
\tocchapterentry {} {References} {260}

}

\vfill\eject

}
%
%
%
%
%
%
\hoffset=0.5truein
\hsize=5.90truein
%
%
\parskip=2pt plus1pt
\parindent=2em
\baselineskip=13pt
%
%
\catcode`\@=11
\topskip=10pt plus 10pt
\r@ggedbottomtrue
\catcode`\@=12
%
%
\font\sslarge=cmss17
\font\bigfancy=eufm24 at 24truept
%
%
\font\eightrm=cmr8
\font\sixrm=cmr6

\font\eighti=cmmi8
\font\sixi=cmmi6
\skewchar\eighti='177 \skewchar\sixi='177

\font\eightsy=cmsy8
\font\sixsy=cmsy6
\skewchar\eightsy='60 \skewchar\sixsy='60

\font\eightbf=cmbx8
\font\sixbf=cmbx6

\font\eighttt=cmtt8

\hyphenchar\tentt=-1 
\hyphenchar\eighttt=-1

\font\eightsl=cmsl8

\font\eightit=cmti8

\newskip\ttglue

\def\eightpoint{\def\rm{\fam0\eightrm}%
  \textfont0=\eightrm \scriptfont0=\sixrm \scriptscriptfont0=\fiverm
  \textfont1=\eighti \scriptfont1=\sixi \scriptscriptfont1=\fivei
  \textfont2=\eightsy \scriptfont2=\sixsy \scriptscriptfont2=\fivesy
  \textfont3=\tenex \scriptfont3=\tenex \scriptscriptfont3=\tenex
  \def\it{\fam\itfam\eightit}%
  \textfont\itfam=\eightit
  \def\sl{\fam\slfam\eightsl}%
  \textfont\slfam=\eightsl
  \def\bf{\fam\bffam\eightbf}%
  \textfont\bffam=\eightbf \scriptfont\bffam=\sixbf
   \scriptscriptfont\bffam=\fivebf
  \def\tt{\fam\ttfam\eighttt}%
  \textfont\ttfam=\eighttt
  \tt \ttglue=.5em plus.25em minus.15em
  \normalbaselineskip=9pt
  \let\sc=\sixrm
  \let\big=\eightbig
  \normalbaselines\rm}
%
%
%
\def\writechaptertoc#1#2{}
\def\writesectiontoc#1#2{}
\def\writesubsectiontoc#1#2{}
%
%
\newcount\chapternum
\newcount\sectionnum
\newcount\subsectionnum
\newcount\eqnum
\newcount\firstpageno
\global\chapternum=0
\global\sectionnum=0
\global\subsectionnum=0
\global\eqnum=0
\def\chapter#1{\global\advance\chapternum by 1\global\eqnum=0
    \global\sectionnum=0\global\subsectionnum=0\vfill\eject
    \centerline{\sslarge Chapter \the\chapternum\ --\ #1}
    \writechaptertoc{\the\chapternum}{#1}
    \global\firstpageno=\pageno
    \def\chaptertitle{#1}
    \edef\savechapterno{\the\chapternum}
    \bigskip\bigskip}
\def\section#1{\global\advance\sectionnum by 1\global\subsectionnum=0
    \goodbreak\bigskip\leftline{\bf \the\chapternum.\the\sectionnum\ #1}
    \writesectiontoc{\the\chapternum.\the\sectionnum}{#1}
    \nobreak\medskip}
\def\subsection#1{\global\advance\subsectionnum by 1 \goodbreak\medskip
    \leftline{\sl \the\chapternum.\the\sectionnum.\the\subsectionnum\ #1}
    \writesubsectiontoc{\the\chapternum.\the\sectionnum.\the\subsectionnum}{#1}
    \nobreak\smallskip}
%
%
\def\quoteskip{2.70truein\relax}
\def\quote#1#2{{\eightpoint\raggedright\pretolerance=10000
    \parskip=1pt\par\noindent{\sl #1}
    \par\hangindent=1.5em\hangafter=0\noindent\llap{--\enspace}#2\par}}
%
%
\def\puteqnum{\global\advance\eqnum by 1
    \eqno{(\the\chapternum.\the\eqnum)}}
%
%
\footline={\ifnum\pageno>0{\hss\tenrm\folio\hss}
           \else{\ifnum\pageno<-2{\hss\tenrm\folio\hss}
                 \else{\hfil}
                 \fi}
           \fi}
%
%
%
\headline={\ifnum\pageno>\firstpageno{\eightpoint\sl \quad
             Chapter \savechapterno\hfil \chaptertitle\quad}
           \else{\hfil}
           \fi}
%
%
%
\def\smallup#1{\raise 1.0ex\hbox{\sixrm #1}}
\def\smallvfootnote#1#2{\vfootnote{{\eightrm\raise 1.0ex\hbox{\sixrm#1}}}
       {\baselineskip=10pt\eightrm #2}}
%
%

\def\figpages#1{\global\advance\pageno by #1}
%
%

%
%
%
%
\def\endchapter{\np}
%
%

\def\ltsima{$\; \buildrel < \over \sim \;$}
\def\ltsim{\lower.5ex\hbox{\ltsima}}
\def\gtsima{$\; \buildrel > \over \sim \;$}
\def\gtsim{\lower.5ex\hbox{\gtsima}}

%
%
\def \K         {{\rm\,K}}
\def \Msun      {{\rm\,M}_{\odot}}
\def \Lsun      {{\rm\,L}_{\odot}}

\def \kms       {{\rm\,km\ s^{-1}}}
\def \mpc       {{\rm\,Mpc}}
\def \yr        {{\rm\,yr}}

\def \cc        {{\rm\,cm^{-3}}}

%
%
\def\xten#1{\times 10^{#1}}
\def\etal{et al{.}\ }
\def\np{\vfill\eject}

\def\eg{e{.}g{.},\ }

\def\numsym{\raise 0.4em\hbox{${\scriptstyle\#}$}}

%
%
%
%
%
%
%
\global\chapternum=0
\global\pageno=1
%
%
%
%

\chapter{Introduction}

%
%

{\leftskip=\quoteskip
\quote{And word by word they handed down the light that shines
today. And those who came at first to scoff, remained behind
to pray.}
{Alan Parsons Project, {\it Ammonia Avenue}}
}

%
%
\bigskip\bigskip

{\bigfancy A} stronomy is a curiosity driven endeavor. Its simple
motivation is captured in the questions ``What is it?'', ``How is it
structured?'', and ``Where did it come from?''. The puzzle is more
enticing because for most observations the only data is the light
transmitted to Earth. One only gets to examine the information carried
in electro-magnetic radiation and is not able to handle, manipulate, or
perform experiments on the objects under study. The intellectual pursuit
must provide the impetus because the knowledge gained serves mainly to
stretch the imagination and has little chance of practical value in
everyday life.  In many ways, it is a science with a philosophical bent:
astronomers are trained to pose questions and search for answers, with
the answers themselves being the reward.

One of the fundamental problems is galaxy formation.  Galaxies are one
of the elemental forms of structure. In grand context, they are the
basic unit of large scale structure in the universe. Closer to home,
galaxies are the characteristic assemblages of smaller structures such
as stars, clusters of stars, interstellar gas, and molecular clouds.  It
is natural to ponder how galaxies formed.

Two viewpoints on the question can be advanced. The first view considers
individual galaxy formation: how does a cloud of matter collapse into
the observed structures of a spiral, elliptical, lenticular, or
irregular galaxy with the corresponding distribution of components? The
second view is concerned with global structure: what cosmological
framework can form galaxies with the observed spatial distribution,
velocity fields, abundances, clustering properties, et cetera? Each form
of the question leads to immense opportunities for study and the problem
of galaxy formation sits at the confluence of these two huge streams of
research.

Each stream is incomplete without the other. It is not sufficient for a
cosmogony to specify only the sites of galaxy formation, it must also
demonstrate that galaxies like those observed can form in those sites.
Likewise, creating beautiful galaxies in isolation becomes meaningful
when one can show that the necessary conditions will arise in the
general universe. It is the goal of this thesis to advance some modest
step along the cross integration of these two streams.

By examining the overlap region, one cannot hope to cover the depth of
either topic. The main concern is for the larger scale aspects of
galaxies and the smaller scale aspects of cosmology. The paragraphs
below will discuss only the topics most relevant to the problem at hand.

\section{Galaxies}

In constructing a theory for the formation of galaxies, the number of
observations one would like to satisfy is immense. The basic shapes of
galaxies are combinations of thin disk and ellipsoidal bulge
components. Disk sizes are typically several hundred pc thick and a few
tens of kpc in diameter while elliptical structures generally range from
kpc to 10 kpc scales. A variety of galaxy eccentricities can be noted
including dwarf objects, cD ellipticals, warps, rings, and tidal
tails. Characteristic masses and luminosities are around $10^{12}\Msun$
and $10^{11}\Lsun$ with several orders of magnitude variation.

Beyond the gross characteristics is the detailed internal structure.
Measurements of radial luminosity profiles show exponential disks and de
Vaucouleur $r^{1/4}$ profiles for elliptical components. The outer
rotation curves of spiral galaxies are flat, indicating a profile of
total mass proportional to radius. Ellipticals are not rotationally
supported, but have their shapes determined by triaxial velocity
dispersions. Significant differences in age and metallicity between
components are also found.

The group properties of galaxies give further constraints. The
luminosity function is fit by a Schechter form, with a power law at low
luminosities and an exponential cutoff at the high end. Spirals dominate
the population in the field, but are found at a reduced fraction in
clustered regions. The rotation velocity of spiral galaxies is
empirically related to luminosity as described by the Tully--Fisher
relation. Similarly for ellipticals, the Faber--Jackson relation
correlates luminosity with central velocity dispersion.

The above paragraphs can only give a flavor for the wide variety of
observations to address. Details and other results can be found in
lengthy reviews of the subject (\eg Silk \& Wyse 1993; Gilmore, Wyse, \&
Kuijken 1989; Efstathiou \& Silk 1983).  The same references plus White
and Frenk (1991) will also serve to expand on the brief review of theory
presented below.

The basic framework for galaxy formation theory is that of gravitational
collapse. In the collisionless case, star formation is assumed to
occur early in the process and the resulting structures are compared with
the old populations found in ellipticals and the bulges of spirals. Though
two body relaxation is negligible for galactic systems, fluctuations in
the potential during collapse leads to a process of violent relaxation
that mixes orbits in phase space (Lynden-Bell 1967). For the final
steady state, triaxial velocity dispersions imply the existence of a
third integral beyond the conservation of energy and angular momentum.
The possibility that galaxy mergers are the dominant process for
elliptical formation has also been raised, but mostly in connection with
simulations (discussed below).

Dissipational collapse is envisioned to form spiral galaxies and many of
the dominant ideas are outlined by White and Rees (1978).
Cooling by radiative processes allows the gas to dissipate thermal
pressure support. The mass scales where cooling is efficient determines
the characteristic mass scale of galaxies (Silk 1977; Rees \& Ostriker
1977). Rotation in the initial cloud slows the collapse in the plane
perpendicular to the angular momentum vector and naturally leads to
flattened disks. A dark collisionless component is assumed in addition
to the gaseous component to account for the flat rotation curves of the
final object. The differences in ages and metallicities of the bulge and
disk populations is accounted for by separate episodes of star
formation; the first occurs during collapse and enriches the metal
content of the interstellar material while the second episode happens
after the disk forms (Eggen, Lynden-Bell, \& Sandage 1962). Assuming
that the dark halo and the gas initially had the same angular momentum
leads to the conclusion that the disks have collapsed by a factor of
order ten relative to the halos (Fall \& Efstathiou 1980). Significant
modifications to this picture can occur when considering the energy
injected from supernova explosions and the effects of a background
ionizing radiation field.

Many tests of these ideas have been pursued via N-body simulations.
Collisionless simulations are able to reproduce elliptical galaxy
profiles with systems that are initially flattened and slowly rotating
(Aarseth \& Binney 1978).  Violent relaxation does not sphericalize the
systems, requiring a third integral. The angular momentum induced by
tidal torques of neighboring perturbations gives low dimensionless spin
parameters of $\lambda\approx 0.07$ with
$\lambda=J|E|^{1/2}G^{-1}M^{-5/2}$ and $J$ is the angular momentum, $E$
is the total energy, $M$ is the mass (Efstathiou \& Jones 1979). Such
values of the spin parameter, when combined with the idea of a dark halo
about spiral galaxies, predict the order of magnitude collapse for disks
and reconcile the collapse time of a disk with the age of the universe
(Fall \& Efstathiou 1980).

The most detailed spiral galaxy formation simulations are those by Katz
and Gunn (1991) and Katz (1992). Their simulations included both
dark matter and baryonic fluids, the physics of gravity, hydrodynamics,
and heating and cooling processes, plus an algorithm to model star
formation with feedback. The simulated galaxies they produced had thin
disks, stellar spheroids, exponential surface mass density profiles, and
flat rotation curves. To the resolution of their model, results were
consistent with the basic scenario outlined above.

Assuming disks are the natural result of a collapse, one might suppose
that ellipticals form by the merging of disk components. A recent review
article by Barnes and Hernquist (1992) covers the details of these
arguments (see also Wielen 1990). The simulations find that the merger
hypothesis generally works well and estimates of the fraction of
ellipticals formed by mergers range from some to most. Incomplete
violent relaxation allows the centers and outskirts of merger remnants
to be composed mostly of particles from the centers and outskirts of the
initial objects, consistent with the metallicity gradients seen in
ellipticals. The merging cross section is governed by the extent of the
dark halos with the luminous parts eventually merging after the halos
have decelerated the systems. Gas in the initial objects is driven
inward by tidal effects and winds up in the core of the merger possibly
providing bursts of star formation or forming small disks. Mergers
provide natural explanations for the slow rotation of giant ellipticals,
double nuclei cD galaxies, and mass shells in some ellipticals. Several
open questions concern the specific number of globular clusters and the
expected core radii in merger remnants.

The main question about these simulations is whether or not the initial
conditions they use are realistic. The distribution of mass before the
collapse is usually assumed to have simple symmetries and one may
question whether such conditions are likely to arise in the general
universe.  For example, the work of Katz and Gunn (1991) assumed a
spherical cloud in solid body rotation and then added small scale
fluctuations to imitate hierarchical effects.  Even with the high
frequency power, the situation is dominated by symmetry and angular
momentum, making disk formation somewhat guaranteed by the physics.
Merging simulations often use models of fully formed galaxies for the
initial objects, although one might expect merging to occur well before
the formation process is complete. One needs to gauge the appropriate
conditions for galaxy formation from large scale models and
establish a connection to cosmology.

\section{Cosmology}

Cosmology is an expansive and quickly changing field. One testament to
its range, and an excellent reference for most of the subjects below, is
the recent 700 page tome by Peebles (1993). Other good general
references include Kolb and Turner (1990) and Peebles (1980). By
comparison, only a meager outline is feasible here.

The same initial conditions problem discussed above is expanded to a
global scale in cosmological studies. Much of the effort in this field
is expended in trying to characterize the nature of the primordial
density fluctuations in the universe. First attempts used optical
surveys, such as the Lick catalog, to estimate on what scale the universe
appeared homogeneous and isotropic. Redshift surveys, which sport many
acronyms like CfA, IRAS, APM, and QDOT, add a third dimension
to better probe the distribution and motions of galaxies.
Large area surveys map the cosmographic features of the
local universe and directed `pencil beam' surveys look for very large
scale properties. Observations of the cosmic microwave background
radiation field, most notably the COBE results (Smoot \etal 1992), confirm
the isotropy of the universe on very large scales and have begun to
provide a baseline measurement of the conditions in the early universe
that can be used to normalize theories. Measurements of the global
density and Hubble parameters constrain the growth of density
fluctuations.

These global concerns are not the focus of this thesis. The more
relevant observations are those closer to galaxy scales. Perhaps the
most fundamental are the correlation function and velocity dispersion of
galaxies in the local universe. These observations set the scale for the
level of clustering and the motions of galaxies. The existence of
quasars at high redshift fixes a minimum timescale in which at least
some objects must collapse. The clouds of the Lyman alpha forest give
clues to the nature and distribution of compact objects back to the era
of quasars. Number counts of faint galaxies and the Butcher--Oemler
effect argue for significant recent evolution of the galaxy
population.

Also pertinent are observations of clusters of galaxies. The dynamical
state of clusters, particularly the presence of substructure, can be
used to estimate the mass scale of objects that have had time to relax
in the universe. The distribution and velocity dispersion of galaxies
can give an estimate of the mass in the cluster. Scaling the mass
estimate by the luminosity density of the universe divided by the
luminosity of the cluster gives an estimate of the global mass density,
with typical values of $\Omega\sim 0.2$ where $\Omega$ is measured in
units of the closure density. X-ray observations show the presence of a
hot intra-cluster medium with a total mass of the same order as that in
the cluster galaxies. Based on this and related cluster data, one may
estimate that baryons comprise about 25\% of cluster mass (White \etal
1993). Recent higher resolution x-ray data indicate the presence of
substructure even in large clusters (White, Briel, \& Henry 1993).

Interpretation of these data begins with gravitational collapse theory.
Starting with a density field in the early universe, one can follow the
evolution of fluctuations in this field. Assuming the initial
fluctuations are small, a perturbation approximation gives the growth in
the linear regime, $\delta\rho/\rho < 1$ where $\delta\rho$ is the
fluctuation and $\rho$ is the mean density. For Gaussian random fields,
the statistics of peaks of the field, which are the expected sites of
galaxy formation, can be considered in detail (Bardeen \etal 1986). A
simpler approximation that predicts the number of collapsed objects of
a given mass scale is given by the Press--Schechter formalism (Press \&
Schechter 1974). Extra freedom is added by noting that galaxies, as peaks
of the distribution, may be more correlated than the underlying mass; an
effect known as biasing.

These tools are applied to a wide variety of structure formation models.
The main element of such models is a description of its mass and energy
components. The fraction in baryons is constrained well by
nucleosynthesis calculations to $\Omega_B\approx 0.0125\,h^2$ where $h$
is the present Hubble parameter in units of $100\kms\mpc^{-1}$. The dark
component comes in two standard flavors, cold and hot, depending on
whether its thermal velocities at the time of matter domination
are non-relativistic or relativistic. The dark matter is assumed to
interact only via gravitation and the amount is governed by one's choice
of $\Omega$ in the model. Inflationary models and the prejudices of
theorists favor $\Omega=1$. Open models assume the lower observational
value and curvature dominated models do both, assuming the observational
value for matter and filling out to unity with a vacuum energy
density. From these assumptions and the class of perturbations
(adiabatic or isocurvature) one generates a spectrum of initial
fluctuations by considering evolution through to the time of decoupling
of baryonic matter and radiation. Alternatively, one may motivate a
fluctuation spectrum from other considerations such as topological
defects or large scale cosmic explosions.

Because galaxies are density enhancements with $\delta\rho / \rho\gg 1$,
linear theory does not address their formation. In probing the
predictions of various cosmological models in the
non-linear regime, N-body simulations have become an indispensable tool.
By following the evolution of representative regions of the universe and
identifying the likely sites for galaxy and cluster formation, one can
`observe' the simulations in similar fashion to real observations and
specify the model predictions. Work
of the so-called ``gang of four'' was influential in shifting focus onto
the cold dark matter (CDM) model (Davis \etal 1985; White \etal 1987b). A
standard CDM model developed with $\Omega=1$, $h=0.5$, and a reasonably
well constrained range of normalizations for the power spectrum.

In recent years, various objections have been raised.  The comparisons
with redshift surveys find more power on large scales than standard CDM
would predict (Maddox \etal 1990; Efstathiou \etal 1990; Vogeley \etal
1992). Large simulations suggest that CDM forms an excessive number of
large mass clusters and requires low normalizations to fit galaxy
velocity dispersions (Gelb 1992).  More definitively, the COBE
observations appear to require a normalization of the CDM power spectrum
on large scales that is much higher than the usual range of
normalizations on galaxy scales.  The current status appears to be that
cold dark matter has been knocked down as king of the hill, but no
successor model has risen to take its place (see Davis \etal 1992; Frenk
1991; Ostriker 1993). Various patches to standard CDM with low $\Omega$,
a cosmological constant, or a mixture of cold and hot dark matter have
been proposed to provide a larger ratio of large scale  to small scale
power that seems to be required (Efstathiou, Sutherland, \& Maddox 1990;
Davis, Summers, and Schlegel 1992; Klypin \etal 1993).

It is important for this work to note that the success of standard CDM
is on the galaxy to cluster scales. The main problem on these scales is
that the inferred velocity dispersions are much higher than those
observed in the local universe. However, enough uncertainties remain
(see the next paragraph) that CDM's strengths outweigh its weaknesses in
this regime. Regardless of which theory supplants CDM, it will have to
be sufficiently CDM-like on the galaxy to cluster scales. Because only
these scales will be addressed, one feels comfortable using CDM as a
representative of hierarchical structure formation for the problem at
hand.

Most of the simulations discussed above assume that the dark matter is
gravitationally dominant and only follow the evolution of a single
dissipationless dark matter fluid. Because baryons are not included,
one can not directly identify where galaxies would form in the
simulations. The unknown bias in the distribution of galaxies versus
the distribution of the dark matter is what allows a range of
normalizations in standard CDM. It has also been suggested that the
velocities of galaxies will be systematically different from those of
the dark matter; an effect dubbed velocity bias. Additionally, objects
composed of collisionless particles quickly lose substructure when
merging and the sites of galaxy formation in cluster regions are washed
out by what is called overmerging. Considering these caveats, there is a
large amount of leeway when interpreting the simulations and falsifying a
cosmogony can be quite difficult. It has sometimes seemed as if CDM was the
theory that would not die.

An important class of simulations has worked to remedy these problems by
including both dark matter and baryonic components. These simulations
have examined the state of the intracluster medium (Evrard 1990; Thomas
\& Couchman 1992) and the intergalactic medium (Ryu, Vishniac, \& Chiang
1990; Cen \& Ostriker 1992a). In cosmology, Cen and Ostriker have
evaluated various scenarios with the two fluid approach, but without the
dynamic range necessary to study galaxy formation. Using a
Lagrangian hydrodynamics code, two groups have been able to resolve
objects with galactic densities within small scale cosmological
simulations (Summers, Davis, \& Evrard 1991; Evrard, Summers, \& Davis
1994; Katz, Hernquist, \& Weinberg 1992). These simulations and the
complex modelling involved are a nascent field, but they offer the
promise of cutting through the biasing fog and being able to identify
galaxies directly.

\section{Galaxies within Cosmology}

The stage is now set for this work. Individual galaxy formation
simulations have progressed to the point where one can produce quite
reasonable galaxy representations on the computer. Large scale structure
investigations have identified a class of models that predict reasonable
sites for galaxy formation. Computational methods have developed the tools
that can begin to handle the severe constraints of the problem. From this
point of view, a thesis such as this one was inevitable.

Although the problem is of huge proportions, the goals of this project
must be more modest. The approach will be to look at the small scale
cosmological simulations (several Mpc) and try to resolve large scale
galaxy phenomena (several kpc). The two fluid, Lagrangian hydrodynamic
simulations will serve as the appropriate vehicle and cold dark matter
will serve as a suitable example of hierarchical structure formation. As
alluded to above, one must first determine whether or not galactic
density objects will form under the conditions of the model. Then one
can check the population of those objects to see if they are
sufficiently ``galaxy-like''. Success in that endeavor allows one to
examine the process of galaxy and cluster formation with confidence that
the results will be useful and applicable when probing galactic and
cosmological questions. The intricate physics involved in galaxy
formation means that one can not hope to model the complete process in
every detail, rather it is prudent to start with the basic physical
processes: gravity, hydrodynamics, shock heating, and radiative cooling.
Further work can include other important processes, such as star
formation and energy feedback from supernovae, and characterize their
effects by comparison to the baseline results.  These investigations
should provide a significant first step toward merging the two streams
of research outlined above.

In addition to the scientific aspects, several computational issues need
to be addressed. The hydrodynamic method that will be used, smoothed
particle hydrodynamics, is being applied to a new regime and its
performance should be tested. A knowledge of how the variations in the
method used by different researchers can affect the results is
imperative for making comparisons. One should also compare the new
results with results one would have obtained by previous methods to show
what consistencies and conflicts arise. Indeed, because the hydrodynamic
methods are computationally very expensive, one wants to justify the
need for their use.  Progress in the computational methods leads to new
avenues of exploration of the science.

The field of galaxy formation is changing rapidly. It is unlikely that
the results presented here will be definitive. The most one can hope for
is that they serve as the current best answer for a brief time, enabling
other research to build on it and progress to a slightly higher level of
understanding yet again. The amount of intellectual effort and just
plain work involved in achieving the modicum of advance presented here
makes one keenly appreciative of the historical groundwork laid down by
so many. The future is sometimes just an elaboration on the past.

%
%

\endchapter

%
%


%
%
%
%
%
\global\chapternum=9
\global\pageno=245
%
%
%
%

\chapter{Summary and Discussion}

%
%

\vbox{\hbox to \hsize{\hfil\pretolerance=300\tolerance=500
\vtop{\hsize=2.65truein
\quote{He's dead, Jim.}{Dr{.} McCoy, {\it Star Trek}}
}\hfil
\vtop{\hsize=2.65truein
\quote{We are able to rule out all $\Omega=1$ cold dark matter models
for $\sigma_8>0.5$ \dots}{Gelb 1992, Ph{.}D{.} Thesis}
}\hfil}}
\bigskip
\vbox{\hbox to \hsize{\hfil\pretolerance=300\tolerance=500
\vtop{\hsize=2.65truein
\quote{He's not dead, he's just resting.}
{Pet Store Owner, {\it Monty Python's Flying Circus}}
}\hfil
\vtop{\hsize=2.65truein
\quote{(these results) are difficult to reconcile with the `standard'
$\Omega=1$, scale-invariant, cold dark matter model $\ldots$}
{Efstathiou \etal 1990, {\it Large-scale Clustering of IRAS Galaxies}}
}\hfil}}
\bigskip
\vbox{\hbox to \hsize{\hfil\pretolerance=300\tolerance=500
\vtop{\hsize=2.65truein
\quote{I'm not dead yet.}{Old man in Scene 2, {\it Monty Python and the
Holy
Grail}}
}\hfil
\vtop{\hsize=2.65truein
\quote{Better understanding of galaxy formation is needed before the
demise of cold dark matter is declared.}
{Davis \etal 1992, {\it The End of Cold Dark Matter?}}
}\hfil}}

%
%
\bigskip\bigskip

{\bigfancy I} t is an idealized view that represents a thesis as an
isolated, compact, and complete piece of research. The simple facade of
collecting work into book form gives the impression that there exists a
beginning, a middle, and an end. Such clean divisions are never so
apparent in reality. A line of study is continuous; inheriting from
previous work and extending into future efforts the ideas, methods, and
knowledge of the present. The boundaries drawn by discrete documents are
somewhat artificial. One might also suppose that when a project can be
truly completed, it will no longer be a forefront enterprise and becomes
an unlikely candidate for a thesis topic.

This work started as a small scale extension to the cosmological dark
matter simulations by the so-called `gang of four' (Davis \etal 1985) as
well as to the hydrodynamic cluster simulations of Evrard (1990). Along
the way, it became apparent that the simulation would really begin to
address galactic scales and thus could make a connection to the isolated
galaxy formation simulations, such as those by Katz (1992). Extensions in
the form of adding star formation and other physics, increasing the
dynamic range, and applying these techniques to larger and smaller scale
problems could now clearly be envisioned. Though many months had been
spent on learning simulation techniques and hundreds of supercomputer
hours had already been burned, it was only at the point when a
continuous path was identified that the project took on a well defined
shape. Crystallization occurred upon establishing context.

This chapter, in combination with the first, is an attempt to define and
describe that context. The pages in between represent another step along
the way, work that shall be collated and correlated here. The various
paths toward future work will be outlined. Individual results take on
both greater and lesser meaning when viewed as part of a larger
progression: greater, because the reason for their significance becomes
apparent, and lesser, because they are viewed as one step in a long
series.

\section{Summary}

This section will have to serve a dual purpose. For those who have read
the previous chapters, it should provide a reminder of the important
results. For those reading only this summary chapter, it should provide
enough depth to foster appreciation of the results. The compromise is
likely to be too heavy in detail for the first group and too thin on
background for the second. The aim is for a proper balance, perhaps a
bit more detailed than a paper abstract, in the capsule chapter
summaries below.

\subsection{Chapter Two}

Chapter 2 presents the simulation methods in detail. The algorithms and
equations of the P3MSPH code are described in a manner geared toward
non-specialists. Emphasis is placed on basic underpinnings of the
methods and upon covering the important elements of all areas of the
code. Special care is taken to motivate aspects of the mesh and SPH
methods, as well as provide connections between them, by discussing an
example in depth. Because SPH techniques are relatively new, some
variations of the choices in the method are explored to promote a feel
for its implementation. The chapter finishes with a discussion of the
limitations of the code due to the necessary numerical compromises.

\subsection{Chapter Three}

A broad overview of the main simulations of this thesis is given in
Chapter 3. The motivations and goals are discussed along with the
specific implementation. Technical parameters are dispensed in enough
detail to allow others to replicate the procedure. As a preemptive
strike against over- and mis-interpretation, limitations of what the
simulation does and does not address are explicitly outlined in advance.
Mass resolution is identified as a critical enabling factor for the
simulation and some related aspects are discussed.

The results of Chapter 3 concern the range of physical regimes addressed
in the simulation. The standard filamentary structures of hierarchical
collapse are found in the 16 Mpc co-moving cubic region, with a large
group of galaxies forming at the vertex of three filaments. The two
particle fluids, representing dark matter and baryonic components, trace
roughly the same structures on large scales, but deviate strongly below
scales of a few hundred kpc. The baryons cool efficiently and, by
radiating their thermal pressure support, collapse to overdensities
characteristic of galaxies. These high density baryonic objects have
small cross sections and do not suffer the excessive merging seen in the
dark matter when clustering to form a group. The dark matter can be
characterized as a single phase fluid peaked at high temperatures and
moderate overdensities, while the baryons exhibit a three phase
structure: a hot component in the halos of groups of galaxies, a cold,
low density component in `voids', and a cold, high density component in
collapsed objects. The simulation aim of covering the smaller
cosmological scales (several Mpc) down to the larger galaxy scales
(several kpc) is confirmed.

\subsection{Chapter Four}

Chapter 4 explores the details of the high density baryonic objects and
their suitability as a simulated galaxy population. The identification
of these galaxy-like objects, or globs for short, is straightforward due
to their high density contrast. In comparison, definition of a dark
matter halo population is less clear cut because of its smooth,
single peaked distribution in density.

The globs and halos have a varied distribution and reasonable
abundances. Of the roughly 200 objects, one third are in groups of two
to nineteen globs, while the rest represent a `field' population. Masses
of the objects range from about 1/30 the Milky Way to a cD-type value
for the largest object of the largest group. The evolution of number
density shows the behavior characteristic of Gaussian random fields and
of hierarchical structure formation. The total mass within collapsed
objects turns out to be several times higher than observations and may
indicate a problem for $\Omega=1$ hierarchical models.  Comparisons of a
glob mass function to the observed galaxy luminosity function are
complicated due to the unknown mass to light ratio, the constrained
overdensity in our simulated region, and necessary final redshift
($z=1$) of the simulation. The faint end mass slope is found to be
$-1.4$, more consistent with a luminosity function corrected for low
surface brightness galaxies ($\approx -1.6$) than with the bright galaxy
data ($\approx -0.95$). The mass distribution within objects, as given
by the circular velocity curves, indicates that the baryons dominate the
central regions while the dark matter governs large scales.

Of special note are the morphologies of the collapsed objects. Numerical
effects will damp velocity dispersions and do not allow the simulation of
elliptical galaxies, halo components, or tidal heating. Galactic disk
structures, linear tidal features, and objects collapsed to unresolved
scales are all predicted and seen. A variety of different structures is
found with more merging and tidal interactions suggested in groups.
Moment of inertia analysis finds disk structures in almost all of the well
resolved, isolated objects. With a condensed baryonic component in the
core, the shapes of the dark matter halos are more spherical in two
fluid simulations than in dark matter only simulations.

Detailed analysis of the disk structures finds several good features.
The sizes, rotation speeds, and the presence of warps of simulated disks
agree well with observed spiral galaxy characteristics. Measured
rotation curves of the particles confirm that the well resolved disks
are rotationally supported. Though pushing the interpretive boundaries,
one does finds a tight Tully-Fisher type relationship and can examine
its consequences.

\subsection{Chapter Five}

In Chapter 5, the formation process of the galactic objects is
investigated. The globs are traced from output to output following the
appearance of new objects, mergers between objects, and dispersal of
objects. The galaxy formation rate, as tracked by the new objects, peaks
at a redshift of $z=4.0$ for a baryonic mass scale of $3\xten{9}\Msun$
and at $z=2.5$ for $10^{11}\Msun$ in baryons.  These early collapse
epochs can easily account for the existence of high redshift quasars,
subject to resolution caveats and the unknown method of powering such
phenomena. Consistent with the expectations of hierarchical structure
formation, the merging rate for these objects is delayed with respect to
the formation rate.  Significant merging also appears correlated with
the formation of the large central group.

The formation sequence for disk objects is followed in detail and a
generic process for hierarchical collapse is identified. Initial
collapse proceeds by the formation of a filament, segmentation of the
filament, and collapse along the filament axis of the local segment.
This one dimensional collapse does not produce significant angular
momentum. Angular momentum is transported into the interior regions of
the forming galaxy by the accretion and dissipation of gaseous clouds.
After that, a large fraction of the final mass is gained by slow
accretion of low density material.

The disk formation process by discrete accretion of gas clouds seems a
natural consequence of hierarchical collapse and differs strongly from
the simple and symmetric analytic picture. While many accretion events
would destroy a disk, the process works because hierarchical formation
requires only a couple significant events to build the next level. To
ensure dissipation, it is important that the clouds remain gaseous
during infall and the timing of star formation could determine the
viability of this process.

With a two fluid simulation, the correspondence between galactic objects
and the dark matter halos can now be pursued. Small mass halos often do
not contain a collapsed baryonic object because the gas has not had
sufficient time to cool to very high overdensities. For well resolved
and isolated systems, there is a good one to one correspondence between
globs and halos. Larger mass globs are found in regions of higher galaxy
density. However, the total mass of globs in group systems is a
decreasing function of halo mass because a larger percentage of the gas
is shock heated to form a hot intra-group medium.

Abundant merging is characteristic of hierarchical formation. The globs
in the group environments generally undergo more merging interactions,
but there is large individual object variance. The dark matter halos see
more mergers than the globs and, for merger events that correlate with
glob mergers, the halo mergers occur about a crossing time or so
earlier. Halo mergers also tend to have larger mass ratios between
objects than found in glob mergers. With a reduced number of mergers and
smaller mass events, disk formation in globs can proceed without too
much disruption. Growth histories of the globs can be used to divine a
nominal mass accretion rate of $10\Msun\yr^{-1}$, but one which is
highly variable due to merging.

\subsection{Chapter Six}

Progressing to larger scales, Chapter 6 looks at how the simulated
galaxies aggregate into groups. In appearance, a striking comparison is
presented between one of the groups in the simulation and a real group
of galaxies from the Palomar plates. Such visual recognition fosters
confidence in the numerical results. For the formation process, a
detailed sequence is traced. The first stage is the collapse of a few
objects in a dense region and their agglomeration to form the core of
the group. Concurrently, the filaments around the group region collapse,
segment, and align toward the core. The next stage is radial
infall of the simulated galaxies from the specific filament directions.
Further accretion will be of groups and small clusters of galaxies, as
higher levels of the hierarchy are assimilated into the dominant
structures.

This formation process has several implications for groups and clusters
in hierarchical scenarios. The prominence of filaments implies
that linear structures should indicate an unvirialized structure in the
infall regime and that projection effects can be severe. Radial orbits
would increase both the relaxation time of the cluster as well as the
frequency of mergers, tidal interactions, and gas stripping from
galaxies.

The distribution of the baryons relative to the dark matter can also be
studied for the group environment. Baryons dominate the mass on galaxy
scales, but relax to cosmic proportions on scales of a few hundred
kiloparsecs. No evidence of segregation of the two fluids is seen on
large scales, suggesting that large baryon fractions in clusters of
galaxies denote problems for $\Omega=1$ models. The cooling of gas into
glob structures is less efficient in larger groups because the
individual galaxy gas halos are ram pressure stripped by the intra-group
medium and heated to the inefficient cooling regime.

Further exploration of the distribution and dynamics of globs within
clusters is needed to address biasing questions. In our small
simulation, the globs are more correlated than the dark matter at high
redshift and less correlated at the final output. A counts-in-cell
analysis indicates that most of the signal for this `anti-bias' arises in
the dense central group region where merging has suppressed clustering
on 1 Mpc scales and increased it on 100 kpc scales.
A scale dependent, but time invariant, velocity bias (the ratio of glob
to dark matter velocity dispersions) is found with a value of 0.7 at the
1 Mpc scale. The mild time dependence of the velocity bias downplays the
role of dynamical friction, but no conclusive source can be identified.
One must be careful not to interpret these results as general
predictions because the simulation models a small and constrained region
of the universe and probably overstates the hydrodynamic effects.

The important comparison with observations is the group mass estimates.
Using the glob population, virial mass estimation of the large group
produces an estimate a factor of two lower than the true binding mass.
Part of this value is due to the small scale clustering of globs and
part to the reduced velocity dispersions. Assuming that luminosity is
proportional to mass in globs, one can scale the group mass estimate to
a global estimate of $\Omega$. Since larger objects form preferentially
in groups, an additional reduction in the estimate occurs and one finds
an example of an $\Omega=0.3$ estimate within an $\Omega=1$ cosmology.
Unfortunately, this example is only a large group that is known to
be unrelaxed and not a rich, virialized cluster.

\subsection{Chapter Seven}

A new direction is taken in Chapter 7. Some of the important processes
one wants to examine are dependent on whether the baryons are gaseous,
and thus dissipational, or stellar, and behave as a
collisionless fluid. A rough parametrization of star formation processes
is introduced to study how the transformation from collisional to
collisionless fluid affects the simulation. Because the resolution
limits are well above real star formation scales, a simple prescription
is employed. The main parameters are  that gas particles are converted
to star particles when their density exceeds $0.1\cc$ and their
temperature is below $3\xten{4}\K$. Thermal energy is injected from
supernovae at a rate of $10^{51}$ ergs per 100 solar masses of star
formation. To maintain a reasonable rate in high density regions and
preserve hydrodynamic resolution, the densest gas particles are turned
into star particles first with nearby candidates being quenched.

The star particles form in the same high density regions that define the
globs. The glob population in the star formation run shows the same
distribution as that in the two fluid run except for some variance in
group regions. The star particles do not remain tightly bound to the
glob overdensities, with only 62\% of the star particles found in globs.
Individual globs range from purely gas particles, in some just collapsed
objects, to purely star particles, with the well resolved globs averaging
a 75\% star particle fraction.  The collisionless distribution of the
star particles smooths out the central peaks of the circular velocity
curves seen in Chapter 4.

The collapse of density peaks proceeds pretty much as in the two fluid
simulation, but the energy feedback of supernovae slows the final stages
and produces a slightly later peak of glob formation, $z=3.5$. Star
particles generally form after the initial collapse of the object and
then continue to be created throughout the life of the glob. The
formation of disks is not disrupted by the presence of star particles
with most of them being converted in the inner 10 kpc of the forming
glob. The discrete accretion scenario for disk formation is roughly the
same because the infalling clouds remain mostly gaseous. Elliptical glob
formation is now possible and a cD-like elliptical forms at the center
of the large group.  A small problem with the quenching process becomes
apparent only for this very large object, due to its high rate of gas
accretion.

The dynamics of the central group are affected in several ways.
First, some small globs that dissipate to the center in the two fluid
run, now survive for much longer and continue their orbits as star
particles. Second, the larger globs have their gas content ram pressure
stripped during radial orbits and proceed on looser, collisionless
orbits. The role of ram pressure in the dynamics of purely gaseous globs
stands in sharp relief. For globs of star particles, it is tidal
dispersion that becomes a strong process, enough so that it is the
standard route through which mergers occur. Finally, the sink of gas
particle mass to star particle mass also acts as a pressure sink in the
inner regions and the overlying structure relaxes inward.

The star formation algorithm generally works well with only minor
modifications necessary at this level of resolution. However,
significant points, such as the relative distributions of stars and gas
within a galaxy, will require marked resolution increases. These
increases will require only straightforward modifications to the
algorithm, but should be accompanied by what appear to be complex
changes in the physics modelled.

\subsection{Chapter Eight}

Chapter 8 considers a computational question directly related to
the results of this thesis: ``How well can one trace galaxies in current
simulations?'' In particular, is the extra modelling and computational
expense of the SPH simulations necessary? A set of criteria that a
suitable galaxy tracer population should meet are defined. Various
methods of tracing sites of galaxy formation are applied to the dark
matter only run, for collisionless algorithms, and the two fluid run,
for hydrodynamic algorithms. Each method is evaluated against the
criteria and then compared against each other.

None of the collisionless algorithms prove satisfactory. The methods
considered are the grouping of particles into halos, the tracking of
peaks in the initial density field, an algorithm proposed by Couchman
and Carlberg (1992), and a new method, dubbed the most bound algorithm,
developed here. The most bound algorithm is a complex attempt to avoid the
problems of the previous methods, but it seems thwarted by the nature of
collisionless interactions. The perturbations in a collisionless fluid
only survive in the central group for a crossing time or so before they
are dispersed by scattering. Dissipation to very large overdensities and
high resolution appear to be required for a collisionless grouping to
maintain coherence.

The SPH methods get passing marks and are reasonably robust. The
definition of globs in this work is compared to the galaxy tracers
defined by Katz \etal (1992) and both methods produce consistent
results. The main arguments against these methods are that the complete
baryonic physics is not included and thus the statistics of the
population could be systematically biased.

When comparing all of these methods on the statistical measures of the
correlation function and velocity dispersion of the galaxy tracers, some
interesting points arise. The collisionless algorithms tend to
undersample the correlation function in cluster regions as noted by
previous workers. An exception is the Couchman \& Carlberg algorithm
which finds an enhanced correlation at all scales, because it follows
only those peaks which have collapsed by an early epoch. Estimates of
the velocity dispersion are anything but straightforward. In particular,
all methods examined would predict a velocity bias in these simulations,
but for very different reasons. Considering the methods and the scale of
simulation used to present velocity bias results, we conclude that no
reliable prediction of velocity bias as a general phenomenon has been
presented.

\subsection{Chapter Nine}

Two tests of the SPH method are presented in Chapter 9. The first
concerns estimating the density field of a particle distribution and how
the various choices in SPH can affect the results. The second is a
dynamical test: the self-similar collapse of an uncompensated top hat
perturbation.

The density tests reveal a lot about resolution in SPH methods. The rms
scatter in density estimates between methods with fixed equivalent
resolution is of order 15\%. When the resolution of methods differs, the
estimates generally get larger with increased resolution. A very small
percentage of particles, $\ltsim 1$\%, can experience order of magnitude
changes in their density estimates when SPH choices are varied. It seems
clear that some estimate of the resolution of various sets of SPH
choices is needed.

The self-similar infall tests provided much information, but only
general conclusions. The reason is that an error in setting a numerical
parameter did not allow the solution to grow properly and prevents
detailed tests of SPH choices.  The results here show that the code
handles the basic gravitational collapse of a collisional gas well and
maintains consistency for large ranges in length, mass, and time scales.
Smoothing and artificial pressure produce only the expected effects.
Different methods for increasing the resolution produce consistent
results. Future work will complete this study.

\section{Discussion}

{}From the above summary of results, it is clear that the basic goals of
this thesis have been met. The model covers sufficient dynamic
range to address galactic scale phenomena within a cosmological scale
simulation. The collapse of density perturbations to overdensities
characteristic of galaxies is followed self-consistently and these
regions cleanly mark the expected sites of galaxy formation. These
results are almost taken for granted throughout much of this work, but
are the key elements upon which the rest is based.

Building on that foundation, one may then examine the simulated galaxy
population. The objects formed are reasonably galaxy-like, with sizes,
masses, and abundances in the range of those for real galaxies.
Concerning morphologies, galactic disks are formed for the first time
from cosmological initial conditions and elliptical objects can be
simulated when star formation is included.  These galactic objects do
not suffer from the excessive merging seen for dark matter halos and can
be followed through clustering. This population of objects forms the
best set to date of galaxy tracers in a cosmological study.

A reliable population allows one to describe the relationship between
the (presumably) luminous baryonic matter and the gravitationally
dominant dark matter. Generally good correspondence is found between the
positions and sizes of galactic objects and dark matter halos, subject
to the local cooling timescale. More data is necessary to quantify
trends of cooled mass and number of collapsed objects versus halo mass.
Circular velocity curves do not indicate an obvious edge to the halos
around galactic objects. These results help develop a feel for the
possible distribution of the unseen component of the universe.

Another opportunity is the study in detail of the hierarchical formation
processes of both individual galaxies and groups of galaxies.  Galactic
disks form via discrete accretion of gas clouds which transport angular
momentum to the inner regions. The collapse of groups of galaxies is
dominated by radial infall along filamentary structures. In
hierarchical structure formation, it must be recognized that infalling
material does not remain static, but will collapse and sub-cluster
during the time of infall.

In addition to the scientific results enumerated above, one also
develops feedback on the methods, ideas, and practices of producing
results. Some points clarify existing questions, some arise
unexpectedly, and others highlight aspects previously not emphasized.
These concerns are often the most valuable ones because they serve to
guide the direction of future work. Several areas of such acquired
knowledge are described below.

\subsection{Methods}

SPH methods of this thesis will continue to be a productive tool for
quite some time. The results of Chapter 8 confirmed that SPH techniques
are both sufficient and necessary for following galaxy formation in
these types of simulations. Standard collisionless methods produce
continual scattering interactions in clustered regions that steadily
disperse sub-perturbations. Carlberg (1993) has suggested that increased
resolution, such that very high overdensities can naturally be resolved
in the collisionless component, may ameliorate the problem. The star
formation runs of Chapter 7 do show less dispersal of star particle
objects when first hydrodynamically collapsed to high density. However,
the radial orbits inherent in the process still produce significant
dispersal of collisionless objects and one might argue the need for
still higher resolution. Even granting that, collisionless methods have
a distinct disadvantage in that the correspondence between halos and
collapsed baryonic objects is dependent on the state of the gas and thus
not easily discernible from dark matter only simulations.

On the other hand, the best formulation of SPH for the problem is not
yet defined. The self-similar spherical infall test of \S9.2 indicates
that the method handles the basic collapse problem well, but only
partial information is available on which parameter choices are optimal.
When feasible, one would favor the fixed method of setting smoothing
lengths because of its stability to variation of kernel and smoothing
formalism as shown in the density tests of \S9.1. Further confidence could
be gained at little cost by combining the fixed method with a density
dependent criteria as described in \S9.1.4. The infall test points out
the need for a dynamical test to gauge the viscosity parameters against
each different set of SPH choices. Although not conclusive, the linear
viscosity parameter of the simulations, $\alpha_1$, should probably be
increased in the future. The choice of smoothing formalisms is found to
be unimportant. No preference is indicated for increased resolution in
either the smoothing kernel or the smoothing lengths, but other
resolution considerations are discussed next.

\subsection{Resolution}

It is identified early on (\S3.5) that sufficient resolution is a key
ingredient in producing many of these results. Later, in the SPH tests,
one notes that the density estimates are dependent on the resolution
even for the same particle configuration. As the cooling of gas scales
as the density squared, one worries that the structures formed will be
controlled by numerical parameter choices. While higher resolution
simulations should see more collapsed structures, it would be unsettling
if one could produce such results by changing only one parameter and not
the full set of related choices.

Two questions arise. First, how does one set a `proper' resolution level
in an SPH simulation? And second, what is the correct measure of
resolution for a set of SPH choices?

The proper resolution level should be directly related to the level of
physical modelling in the simulation. One must identify the range of
structures one hopes to resolve, but only those for which the
relevant physical processes are included. These structures should be
related to the length scales of the gravitational softening and the
hydrodynamic smoothing. The mass resolution should
consider the total mass of a resolved object, about 30 particles for
SPH. The relevant temporal criteria are the dynamical and crossing times
of the structures. Above all, these parameters must form a consistent
set and should not be viewed as a bunch of numerical dials that can be
tuned independently.

As an example, one can check the main simulation of this thesis. The
mass per baryonic particle is about $10^{8}\Msun$, giving a total mass
in a minimally resolved structure of about $3\xten{10}\Msun$. Such
structure is modelled through collapse only down to a temperature of
$10^{4}\K$ and sub-galactic structure is not handled. Minimum values of
7 kpc for the gravitational softening and 3.5 kpc for the hydrodynamic
smoothing produce an appropriate 10 kpc length resolution. The timestep
of 6 Myr ensures that a velocity of 500 km/s covers only a third of a
length scale in one timestep. Velocity dispersions of small objects are
much lower, but relative velocities in large potential wells are of this
order. One element that is not modelled correctly is that the cooling
timescales can be very short, but this occurs only for high density
objects and does not pose much of a problem.

The SPH choices, unfortunately, do not yet have a quantitative measure
of their resolution. Perhaps a standardized density test could be
devised. A more relevant test for the galaxy formation problem might be
one that checks at what level cooling is resolved. For our choices of
SPH parameters, we find that significant cooling is resolved with a
minimum resolved mass near or below $10^{12}\Msun$. This figure agrees
with CDM expectations (\eg Blumenthal \etal 1984) and
indicates SPH resolution consistent with the 30 particle level. Sharper
kernels and smaller smoothing lengths may resolve the hydrodynamics at
an inconsistent level.

\subsection{Physics}

One thing a simulationist must always recognize is that
even the best numerical effort will contain an incomplete physical model.
The results will always be subject to caveats. Section 3.2 is an attempt
to outline the major caveats in advance for the simulations of this
thesis. In this section, it will serve to examine the processes behind
those caveats.

Chapter 7 provides a short investigation of the effects of star
formation. More accurately, the consideration is for what happens upon
turning the gas particles into collisionless particles than for the
impossible feat of including a realistic star formation model. The
results point out that ram pressure effects may be overstated in the two
fluid model. Though one can now follow galaxy evolution internal to
clusters, the dynamics and distribution are not yet fully reliable.
Star formation runs will have to be further studied to get a handle on
the dispersal rates of galactic objects and the relative strengths of ram
pressure and dynamical friction.

Further modelling of star formation will require only straightforward
changes to the simple model of Chapter 7. The one problem that surfaces
can be fixed without trouble. The ideas presented by Katz (1992) and
Navarro and White (1993) can be implemented when the resolution justifies
their inclusion and should suffice in the foreseeable future. Although we
are a long way from molecular cloud scales, some thought may have to be
given to the star formation in globular clusters. Here we note the
reciprocal of the advice of the previous section, the modelling should
be geared to the resolution scale.

Another relevant process is the effects of a background radiation field.
The proximity effect of Lyman alpha clouds near quasars indicates that
a substantial amount of background ionizing radiation exists in the
early universe (Bechtold 1993). This radiation would have no
heating effect on the current simulation because temperatures below
$10^4\K$ are not modelled. The major change would be in shifting the
gas away from collisional ionization equilibrium and significantly
altering the cooling curve. Adding photoionization will decrease
the rate of collisional ionizations and drop the amount of thermal
energy radiated. While large objects would attain sufficient densities
to shrug off these effects, small perturbations may cool much less
efficiently. The era of galaxy formation would be expected to be
delayed.

The most difficult physical elements to include involve the processes
below a temperature of $10^4\K$. A complex set of atomic and molecular
radiation processes are necessary to cool gas to much lower
temperatures. The extra variables necessary to store species fractions
will create significant memory problems. The timescales involved are
much shorter than the simulation timesteps and produce stiff equations.
In addition, one needs to model a process of cooling for primordial gas
and then keep track of metallicity from supernova enrichment. These
prospects seem challenging, but realistic modelling of the internal
structure of galaxies is dependent on it.

\section{Future Paths}

This thesis has emphasized the direct results from computer simulations,
but they are not an end in themselves. The best role of simulations is
to provide insight and guide one's intuition in developing theoretical
understanding. The natural follow-on to this work is to re-examine the
formation theories in the light of the processes identified here. The
idea of creating galactic disks through discrete accretion of gas clouds
will have to be considered in detail to see if it can produce the
standard features like thin and thick disks, warps, and the observed
radial gas distribution, as well as the unusual characteristics such as
counter-rotating or bi-directional stellar disks. One should also pursue
the implications of filament dominated infall for cluster formation and
interpretation of those observations. These studies will provide a
significant evaluation of hierarchical structure formation.

Further simulation work can delve into the regions not adequately
covered here. The most important extension is a simulation of an
unconstrained region of space evolved to the present day. The processes
identified in this work should be present in a lower density volume, but
may be delayed and proceed more slowly. Because the current simulation
ends at approximately one third of the age of the universe, the
stability of structures and the merging histories can be followed for
considerably longer. Comparisons between the numerical results and
present day observations will be facilitated.

One would also like to explore processes on other scales. Larger volumes
will consider more levels of the hierarchy and will address cosmological
concerns better. Smaller volumes can study the high resolution formation
of a few objects. Each scale will have the attendant concerns about
resolution and physical modelling.

Star formation algorithms will be employed where appropriate, but
emphasis should be placed on studies of clusters of galaxies. In
particular, it is very important to understand the relative roles of ram
pressure, dynamical friction, and tidal disruption in order to make
predictions about the possible biases between the distributions of the
galaxies, the x-ray gas, and the dominant mass. These factors enter
heavily in the interpretation of cluster observations and their
implications for global structure.

A final scientific point to address is the effects of varying the
cosmological model. Recent observational results have led to the general
belief that cold dark matter requires some adjustments, but no single
model has emerged as a favorite. On the scale of these simulations, one
can check such things as the timing of galaxy formation and the
variations in the formation processes for mixed dark matter or low
$\Omega$ models. Most work on these models has been done with poor
resolution of galaxy scales and these questions become very important
for the viability of the models.

Modifications to the computer code will follow the guidelines of the
previous section. Increases in resolution and physical modelling will
progress together. Investigations of including a background ionizing
radiation field have been started by Weinberg (private communication)
while detailed radiation processes have been explored by Cen (1992).

To accommodate extended dynamic range and modelling, one will also have
to increase the speed and efficiency of the code. In work on adapting a
P3M code to parallel machines, Ferrell \& Bertschinger (1993) have
parallelized the particle mesh part in generalized fashion, but
Bertschinger (private communication) finds that the direct summation
implementation varies in method and efficiency among different machines.
SPH will introduce some constraints on the algorithms because one must
ensure that a direct summation is performed at each particle over a
set minimum number of neighbor particles. Parallel architectures have
historically been a shifting target, but the development of High
Performance FORTRAN and the MPP FORTRAN programming model (Pase \etal
1993) are leading toward standardization. Another way to attack the problem
is through specialized hardware. The GRAPE processor boards (Ito
\etal 1991) perform a direct summation gravity calculation in hardware
and accumulate lists of neighboring particles for SPH calculations.  The
current production version has a Plummer force law hardwired and is well
suited to a combination Aarseth and SPH code (Umemura \etal 1993).  We
are investigating whether it can be adapted to the combination of force
laws needed in P3M codes. Production versions of GRAPE boards with a
programmable force law are expected in the next year or so.

The tests of the SPH code in Chapter 9 should be continued and
augmented. The self-similar infall tests are now at a point where
quantitative tests of parameter choices can be done. The relationship
between density estimates, resolution, and cooling needs to be explored.
The first approach would be a well defined problem, such as estimating
the density in a power law distribution. After that, a series of small
cosmological
simulations with varying SPH choices could provide direct feedback on the
processes relevant to galaxy formation studies.

The projects outlined above will follow through on the context
envisioned in the introduction of this chapter. The flow of research
should always progress through the current state toward future goals.
The importance of this work lies as much in the new paths to which it
leads, as in the old questions it answers.

\section{Some Final Remarks}

At the culmination of any long and involved journey, one is oft given to
philosophizing. The broad exposition of events, details, and results
yields to a more personal and introspective interpretation. Even when
the presumed goal is rational, objective, and quantifiable scientific
knowledge, the essence of being human exerts itself in the need to pass
final judgement based on a purely subjective measure of individual
enrichment. Permit me a small indulgence of these desires.

I hold some very different views now than when I started my thesis work.
I began by imagining how grand it would be to follow the collapse of
galactic density perturbations, to really visualize the collapse
process, and to gain intuition for the details and subtleties.  The
prospect of becoming expert in simulation codes and producing realistic
models was exciting indeed. And even with wide-eyed ideas, the project
still surpassed my expectations. But now, the goals have shifted their
stature.

Familiarity and intimate knowledge are without peer in granting
perspective. Within the global context of a rich field of research and
the local context of assumptions and caveats, lofty ideals are softened.
My original aims are now viewed in their relative merit to the progress
and prospects of the field. Enthusiasm is undamped, but I focus on
exploring new regimes, rather than achieving specific targets. My hopes
for the future are simply that it lives up to the promise of the past.

And thus, in a personal sense, this thesis is not so much the story of
scientific research as it is the story of the person doing that
work: a rite of passage from inexperienced graduate student, learning
from those gone before, to (somewhat) knowledgeable researcher, able to
join one's peers in disseminating new information.  The personal journey
is the most valuable lesson, and one which cannot be embodied in a
thesis or granted by a Ph{.}~D. diploma. It never ends. We shall always
be students to experience.

%
%

\endchapter

%
%


%
%
%
%
\global\pageno=260
%
%
\def\refpar{\par\noindent\hangindent=1.5em\frenchspacing}
\def\bref{\refpar}
\def\jref#1;#2;#3;#4;#5;#6.{\refpar #1 #2, {\it #3} {\bf #4}, #5.}
\def\jrefpreprint#1;#2;#3;#4;#5;#6.{\refpar #1 #2, preprint.}
\def\jrefsub#1;#2;#3;#4;#5;#6.{\refpar #1 #2, {\it #3}, submitted.}
\def\jrefpress#1;#2;#3;#4;#5;#6.{\refpar #1 #2, {\it #3}, in press.}
%
%
\def\ana{Astr. Astrophys.}
\def\aj{Astr. J.}
\def\apj{Astrophys. J.}
\def\apjl{Astrophys. J.}
\def\apjs{Astrophys. J. Suppl. Ser.}
\def\araa{A. Rev. Astr. Astrophys.}
\def\baas{Bull. Am. Astr. Soc.}
\def\cip{Comput. Phys.}
\def\fcp{Fund. Cosmic Phys.}
\def\ijmpc{Intl. J. Mod. Phys. C}
\def\jcp{J. Comp. Phys.}
\def\mnras{Mon. Not. R. astr. Soc.}
\def\nature{Nature}
\def\pasj{Pub. Astr. Soc. Japan}
\def\physrpts{Phys. Rpts.}
\def\physicascripta{Physica Scripta}
\def\siamjssc{SIAM J. Sci. Stat. Comput.}
\def\revmodphys{Rev. mod. Phys.}
%
%
\vfill\eject
%
%
\global\firstpageno=\pageno
\headline={\ifnum\pageno>\firstpageno{\eightpoint\sl\hfil
             References\quad}
           \else{\hfil}
           \fi}
%
%
%
%
%
%

\centerline{\sslarge References}
\bigskip\bigskip

\jref
Aarseth, S. J., \& Binney, J.;1978;\mnras;185;227;.

\jref
Arnaud, M., Rothenflug, R., Boulade, O., Vigroux, L., \&
Vangioni--Flam, E.;1991;\ana;254;49;.

\jref
Babul, A., \& White, S. D. M.;1991;\mnras;253;31p;34p.

\jref
Bardeen, J. M., Bond, J. R., Kaiser, N., \& Szalay, A. S.;1986;\apj;304;15;.

\jref
Barnes, J.;1989;\nature;338;123;.

\jref
Barnes, J. E., \& Hernquist, L.;1992;\araa;30;705;742.

\jref
Bean, A. J., Efstathiou, G., Ellis, R. S., Peterson, B. A., \& Shanks,
T.;1983;\mnras;205;605;.

\jrefpress
Bechtold, J.;1993;\apjs;;;.

\jref
Bertschinger, E.;1985;\apjs;58;39;66.

\jref
Bertschinger, E.;1987;\apjl;323;L103;.

\jref
Bertschinger, E., \& Gelb, J. M.;1991;\cip;Mar/Apr;164;179.

\jrefsub
Bertschinger, E., \& Jain, B.;1993;\apj;;;.

\jref
Bicknell, G. V.;1991;\siamjssc;12;1198;1206.

\jref
Bingelli, B., Sandage, A., \& Tammann, G. A.;1985;\aj;90;1681;.

\jref
Blanchard, A., Valls--Gabaud, D., \& Mamon, G. A.;1992;\ana;264;365;.

\jref
Blumenthal, G. R., Faber, S. M., Primack, J. R., \& Rees, M. J.;1984;\nature
;311;517;525.

\jref
Bothun, G. D., Impey, C. D., \& Malin, D. F.;1991;\apj;376;404;.

\bref
Boyle, B. J. 1993, in {\it The Evolution of Galaxies and Their
Environment}, eds. H. A. Thronson \& J. M. Shull, in press.

\jref
Broadhurst, T. J., Ellis, R. S., \& Shanks, T.;1988;\mnras;235;827;.

\jref
Carignan, C., \& Freeman, K. C.;1988;\apjl;332;L33;.

\jref
Carlberg, R. G.;1988;\apj;324;664;.

\jref
Carlberg, R. G.;1990;\apjl;359;L1;.

\jref
Carlberg, R. G.;1991;\apj;367;385;.

\jrefpreprint
Carlberg, R. G.;1993;;;;.

\jref
Carlberg, R. G., \& Couchman, H. M. P.;1989;\apj;340;47;.

\jref
Carlberg, R. G., Couchman, H. M. P., \& Thomas, P. A.;1990;\apjl;352;L29;L32.

\jref
Carlberg, R. G., \& Dubinski, J.;1991;\apj;369;13;.

\jref
Casertano, S., \& Hut, P.;1985;\apj;298;80;.

\jref
Cavaliere, A., Colafrancesco, S., \& Scaramella, R.;1991;\apj;380;15;23.

\jref
Cen, R.;1992;\apjs;78;341;.

\jref
Cen, R., \& Ostriker, J.;1992a;\apj;393;22;41.

\jref
Cen, R., \& Ostriker, J.;1992b;\apjl;399;L113;L116.

\jref
Cole, S.;1991;\apj;367;45;53.

\jref
Cole, S., \& Kaiser, N.;1989;\mnras;237;1127;.

\jref
Colless, M. M., Ellis, R. S., Taylor, K., \& Hook, R. N.;1990;\mnras
;244;408;423.

\jref
Couchman, H. M. P.;1991;\apj;368;L23;.

\jref
Couchman, H. M. P., \& Carlberg, R. G.;1992;\apj;389;453;463.

\jref
David, L. P., Arnaud, K. A., Forman, W., \& Jones, C.;1990;\apj
;356;72;.

\jref
Davis, M., Efstathiou, G., Frenk, C. S., \& White, S. D. M.;1985;\apj
;292;371;394.

\jref
Davis, M., Efstathiou, G., Frenk, C. S., \& White, S. D. M.;1992;\nature
;356;489;494.

\jref
Davis, M., \& Peebles, P. J. E.;1983;\apj;267;465;.

\jref
Davis, M., Summers, F\thinspace J, \& Schlegel, D.;1992;\nature;359;393;396.

\jref
Dekel, A., \& Silk, J.;1986;\apj;303;39;.

\bref
Djorgovski, S., De Carvalho, R. \& Han, M.-S. 1988, in {\it Extragalactic
Distance Scale}, ed. S. van den Bergh, ASP Conf. Ser., 4, 203.

\jref
Dressler, A.;1980;\apj;236;351;.

\jref
Dubinski, J., \& Carlberg, R.;1991;\apj;378;496;.

\jref
Efstathiou, G.;1992;\mnras;256;43p;.

\jref
Efstathiou, G., Bond, J. R., \& White, S. D. M.;1992;\mnras;258;1p;6p.

\jref
Efstathiou, G., Davis, M., Frenk, C. S., \& White, S. D. M.;1985;\apjs
;57;241;260.

\jref
Efstathiou, G., Frenk, C. S., White, S. D. M., \& Davis, M.;1988;\mnras
;235;715;748.

\jref
Efstathiou, G., \& Jones, B. J. T.;1979;\mnras;186;133;.

\jref
Efstathiou, G., Kaiser, N., Saunders, W., Lawrence, A., Rowan-Robinson,
M., Ellis, R. S., \& Frenk, C. S.;1990;\mnras;247;10p;14p.

\jref
Efstathiou, G., \& Rees, M. J.;1988;\mnras;230;5p;.

\jref
Efstathiou, G., \& Silk, J.;1983;\fcp;9;1;138.

\jref
Efstathiou, G., Sutherland, W. J., \& Maddox, S. J.;1990;\nature
;348;705;707.

\jref
Eggen, O. J., Lynden-Bell, D., \& Sandage, A. R.;1962;\apj;136;748;766.

\jref
Evrard, A. E.;1987;\apj;316;36;.

\jref
Evrard, A. E.;1988;\mnras;235;911;.

\jref
Evrard, A. E.;1990;\apj;363;349;366.

\jrefpress
Evrard, A. E., Summers, F\thinspace J, \& Davis, M.;1994;\apj;;;.

\bref
Faber, S. M. 1982, in {\it Astrophysical Cosmology}, eds. H. A. Bruck,
G. V. Coyne, and M. S. Longair, Pontificia Academia Scientiarium,
Vatican, 191.

\jref
Fall, S. M., \& Efstathiou, G.;1980;\mnras;193;189;.

\jref
Fall, S. M., \& Rees, M. J.;1985;\apj;298;18;.

\jrefsub
Ferrel, R., \& Bertschinger, E.;1993;\ijmpc;;;.

\jref
Frenk, C. S.;1991;\physicascripta;T36;70;87.

\jref
Frenk, C. S., White, S. D. M., Davis, M., \& Efstathiou, G.;1988;\apj
;327;507;525.

\bref
Gelb, J. M. 1992, Ph{.}D. Thesis, MIT.

\jref
Gilmore, G., Wyse, R. F. G., \& Kuijken, K.;1989;\araa;27;555;627.

\jref
Gott, J. R., \& Thuan, T. X.;1976;\apj;204;649;.

\jref
Heisler, J., Tremaine, S., \& Bahcall, J. N.;1985;\apj;298;8;17.

\jref
Hernquist, L., Bouchet, F. R., \& Suto, Y.;1991;\apjs;75;231;240.

\jref
Hernquist, L., \& Katz, N.;1989;\apjs;70;419;.

\bref
Hockney, R. W., \& Eastwood, J. W. 1988, {\it Computer Simulation Using
Particles}, Adam Hilger, New York.

\jref
Impey, C., Bothun, G., \& Malin, D.;1988;\apj;330;634;.

\jref
Ito, T., Ebisuzaki, T., Makino, J., \& Sugimoto, D.;1991;\pasj
;43;547;555.

\jref
Katz, N.;1992;\apj;391;502;517.

\jref
Katz, N., \& Gunn, J. E.;1991;\apj;377;365;381.

\jref
Katz, N., Hernquist, L., \& Weinberg, D. H.;1992;\apjl;399;L109;L112.

\jrefsub
Katz, N., Quinn, T., \& Gelb, J. M.;1993;\mnras;;;.

\jref
Kaiser, N.;1986;\mnras;222;323;.

\bref
Kennicut, R. C. 1993, in {\it The Evolution of Galaxies and Their
Environment}, eds. H. A. Thronson \& J. M. Shull, in press.

\jref
Kirshner, R. P., Oemler,  A., Schechter, P. L., \& Schectman, S. A.;1983;\aj
;88;1285;.

\jref
Klypin, A., Holtzman, J., Primack, J., \& Reg\"os, E.;1993;\apj;416;1;16.

\bref
Kolb, E. W., \& Turner, M. S. 1990, {\it The Early Universe}, Addison-Wesley,
Redwood City, CA.

\jref
Kriss, G. A., Cioffi, D. F., \& Canizares, C. R.;1983;\apj;272;439;.

\jref
Larson, R. B., Tinsley, B. M., and Caldwell, C. N.;1980;\apj;237;692;.

\jref
Lattanzio, J. C., Monaghan, J. J., Pongracic, H., \& Schwarz,
M. P.;1986;\siamjssc;7;591;598.

\jref
Lilly, S. J., Cowie, L. L., \& Gardner, J. P.;1991;\apj;369;79;.

\jref
Lin, D. N. C., \& Lynden-Bell, D.;1982;\mnras;198;707;.

\jref
Lin, D. N. C., \& Murray, S. D.;1992;\apj;394;523;533.

\jref
Loveday, J.,  Peterson, B. A., Efstathiou, G., \& Maddox, S. J.;1992;\apj
;390;338;.

\jref
Lubin, L. M., \& Bahcall, N. A.;1993;\apjl;415;L17;L20.

\jref
Lynden-Bell, D.;1967;\mnras;136;101;.

\jref
Maddox, S. J., Efstathiou, G., Sutherland, W. J., \& Loveday,
J.;1990;\mnras;242;43p;47p.

\jref
Miralda--Escud\'e, J., \& Ostriker, J. P.;1990;\apj;350;1;22.

\jref
Monaghan, J. J.;1982;\siamjssc;3;422;433.

\jref
Monaghan, J. J.;1992;\araa;30;543;574.

\jref
Monaghan, J. J., \& Gingold, R. A.;1983;\jcp;52;374;389.

\jref
Monaghan, J. J., \& Lattanzio, J. C.;1985;\ana;149;135;143.

\jref
Navarro, J. F., \& Benz, W.;1991;\apj;380;320;.

\jrefpreprint
Navarro, J. F., \& White, S. D. M.;1993;;;;.

\jref
Nolthenius, R., \& White, S. D. M.;1987;\mnras;225;505;530.

\jref
Ostriker, J. P.;1993;\araa;31;689;716.

\jref
Park, C.;1991;\mnras;251;167;173.

\bref
Pase, D. M., MacDonald, T., \& Meltzer, A. 1993, {\it MPP Fortran
Programming Model}, Cray Research, Available via anonymous ftp from
ftp.cray.com.

\bref
Peebles, P. J. E. 1980, {\it The Large Scale Structure of the Universe},
Princeton University Press, Princeton, NJ.

\bref
Peebles, P. J. E. 1993, {\it Principles of Physical Cosmology}, Princeton
University Press, Princeton, NJ.

\jref
Persic, M., \& Salucci, P.;1988;\mnras;234;131;.

\jref
Persic, M., \& Salucci, P.;1992;\mnras;258;14p;18p.

\jref
Phillipps, S., \& Disney, M.;1985;\ana;148;234;.

\jref
Postman, M., \& Geller, M. J.;1984;\apj;281;95;.

\jref
Press, W. H., \& Schechter, P.;1974;\apj;187;425;.

\jref
Rees, M. J., \& Ostriker, J. P.;1977;\apj;179;541;.

\jref
Rubin, V. C., Burstein, D., Ford, W. K., \& Thonnard, N.;1985;\apj
;289;81;104.

\jref
Ryu, D., Vishniac, E. T., \& Chiang, W. H.;1990;\apj;354;389;.

\jref
Sandage, A., Tammann, G. A., \& Yahil, A.;1979;\apj;232;352;.

\jref
Sarazin, C. L.;1986;\revmodphys;58;1;.

\bref
Shanks, T., Fong, R., Boyle, B. J., \& Peterson, B. A. 1989,
in {\it The Epoch of Galaxy Formation}, eds. C. S. Frenk \etal
Kluwer Academic, Dordrecht, 141.

\jref
Silk, J. I.;1977;\apj;211;638;.

\jref
Silk, J. I., \& Wyse, R.;1993;\physrpts;231;293;365.

\jref
Smoot, G. F., \etal;1992;\apjl;396;L1;.

\jref
Summers, F\thinspace J, Davis, M., \& Evrard, A. E.;1991;\baas;23;1343;.

\jref
Thomas, P. A., \& Couchman, H. M. P.;1992;\mnras;257;11;31.

\jref
Toomre, A., \& Toomre, J.;\apj;1972;178;623;.

\jref
T\'oth, G., \& Ostriker, J. P.;1992;\apj;389;5;26.

\jref
Umemura, M., Fukushige, T., Makino, J., Ebisuzaki, T., Sugimoto, D.,
Turner, E. L., \& Loeb, A.;1993;\pasj;45;311;320.

\jref
Vogeley, M. S., Park, C., Geller, M. J., \& Huchra, J. P.;1992;\apjl
;391;L5;L8.

\jref
Walker, T. P., Steigman, G., Schramm, D. N., Olive, K. A.,
\& Kang, H.;1991;\apj;376;51;69.

\jref
Weinberg, D. H.;1992;\mnras;254;315;342.

\jref
White, S. D. M., Briel, U. G., \& Henry, J. P.;1993;\mnras;261;L8;L12.

\jref
White, S. D. M., Davis, M., Efstathiou, G., \& Frenk, C. S.;1987a;\nature
;330;451;453.

\jref
White, S. D. M. \& Frenk, C. S.;1991;\apj;379;52;79.

\jref
White, S. D. M., Frenk, C. S., \& Davis, M.;1983;\apjl;274;L1;L5.

\jref
White, S. D. M., Frenk, C. S., Davis, M., \& Efstathiou, G.;1987b;\apj
;313;505;516.

\jrefsub
White, S. D. M., Navarro, J. F., Evrard, A. E., \& Frenk, C.
S.;1993;\nature;;;.

\jref
White, S. D. M. \& Rees, M.;1978;\mnras;183;341;358.

\bref
Wielen, R., ed. 1990, {\it Dynamics and Interactions of Galaxies},
Springer-Verlag, Berlin.

\jref
Zurek, W., Quinn, P., \& Salmon, J.;1988;\apj;330;519;.

%
%
\vfill\eject
%
%
%
%
%
%


%
%
\end